\documentclass[runningheads]{svmult}

\usepackage{makeidx}   
\usepackage{graphicx}  
\usepackage{subeqnar}  
\usepackage{multicol}  
\usepackage{physprbb}  
\makeindex             

\begin{document}
\title*{~~~~~~~~~~~~~~~~~~~ The TESIS project \protect\newline 
Are type 2 QSO hidden in X-ray emitting EROs?}
\toctitle{The TESIS project:\protect\newline
Are type 2 QSO hidden in X-ray emitting EROs?}
%
%
\titlerunning{Are type 2 QSO hidden in X-ray emitting EROs?}
%
\author{P. Severgnini\inst{1}
\and R. Della Ceca\inst{1}
\and V. Braito\inst{1}
\and P. Saracco\inst{1}
\and M. Longhetti\inst{1}
\and R. Bender\inst{2}
\and N. Drory\inst{3}
\and G. Feulner\inst{2}
\and U. Hopp\inst{2}
\and F. Mannucci\inst{4}
\and C. Maraston\inst{5}
}
\authorrunning{P. Severgnini et al.}
%
%
\institute{INAF-Osservatorio Astronomico di Brera (OAB), via Brera 28, 20121 Milano, Italy
\and Universit\"ats-Sternwarte M\"unchen, Scheiner Str. 1, 81679 M-unchen, Germany
\and University of Texas at Austin, Austin, Texas 78712
\and IRA-CNR, Largo E. Fermi 5, 50125 Firenze, Italy
\and Max-Plank-Institut f\"ur Extraterrestrische Physik, Garching bei Munchen, Germany
}

\maketitle              



\section{Introduction}
X-ray selected EROs are, on average, the hardest X-ray
sources in medium and deep X-ray fields. This coupled with their extremely
red colors (R-K$>$5) suggest that they represent one of the most promising population
where looking for high-luminosity (L$_X$$>$10$^{44}$ erg s$^{-1}$) and X-ray obscured 
(N$_H>$10$^{22}$ cm$^{-2}$) type2 AGNs, the so called QSO2 
(e.g. \cite{vig}; \cite{stev}; Mignoli et al. submitted to A\&A). These latter are predicted in large density by the 
synthesis model of the Cosmic X-ray background (\cite{gil}) even if only 
few observational evidences have been found so far 
(e.g. \cite{del} and references therein; Caccianiga et al. A\&A accepted).

With the aim at studying the properties of X-ray emitting EROs
our group has recently gained 75 ksec (AO-2 period) and 78 ksec
(AO-3 period) of XMM-Newton observations
centered on two  MUNICS  (MUNICS Near-IR Cluster Survey, \cite{dro})
fields of about 180 arcmin$^2$  each. For these fields 
photometric observations in the B, V, R, I, J, K'
bands are available (Drory et al. 2001) down to limiting magnitudes of R$\sim$24.5
mag and K'$\sim$19.5 mag. All the EROs present in these 2 fields having a K' magnitude
brighter than 18.5 are already under investigation by the TESIS (TNG EROs Spectroscopic
Identification Survey, Saracco et al. this conference and references therein) project.

Here we present the analysis of the XMM-Newton data for the S2F1 field and we
discuss the X-ray properties of the 6 X-ray emitting EROs detected down to 
$\sim$10$^{-15}$ erg cm$^2$ s$^{-1}$. 
We assume H$_0$=65 km s$^{-1}$ Mpc$^{-1}$ and q$_0$=0.

\section{X--ray data analysis}
The field S2F1 was observed by XMM-Newton on February 11, 2003.
The data have been cleaned and processed using the Science Analysis System
version 5.3.3. The net exposure time after data cleaning is about 41 ksec.
We have cross--correlated the astrometrically corrected X–-ray source catalogue
with the TESIS-EROs catalogue. We found 5 secure
X-ray emitting EROs and one more X-ray source which is more likely associated with
an ERO counterpart. The detailed analysis of the 6 sources is presented
in Severgnini et al. (in preparation).
On the basis of their photometric redshifts all of the 6 X-ray emitting EROs
have observed 2-10 keV luminosities (L$_{(2-10~keV)}>$10$^{42}$ erg s$^{-1}$) and 
X--ray--to--optical flux ratios (Log(F$_{(2-10~keV)}$)/Log(F$_R)>$-1) 
suggesting the presence of AGN.
For three out of the 6 EROs we  have been able to perform a
complete X-ray spectral analysis.
The X-ray data are well fitted by a single power-law model (Fig.~1).
\begin{figure}[h!]
\vskip -0.3truecm
\begin{center}
\includegraphics[width=4cm, height=5cm, angle=-90]{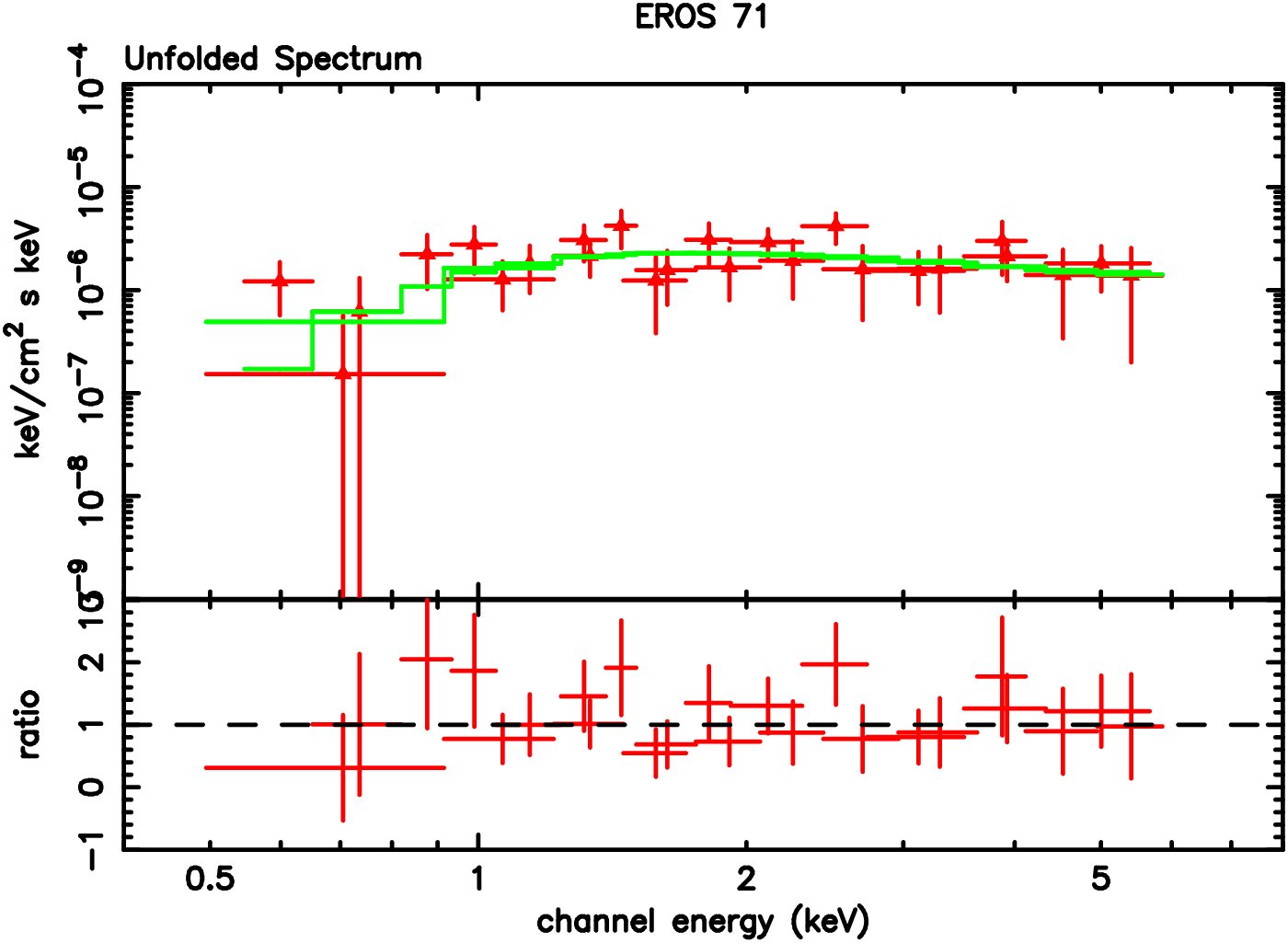}
\includegraphics[width=4cm, height=5cm, angle=-90]{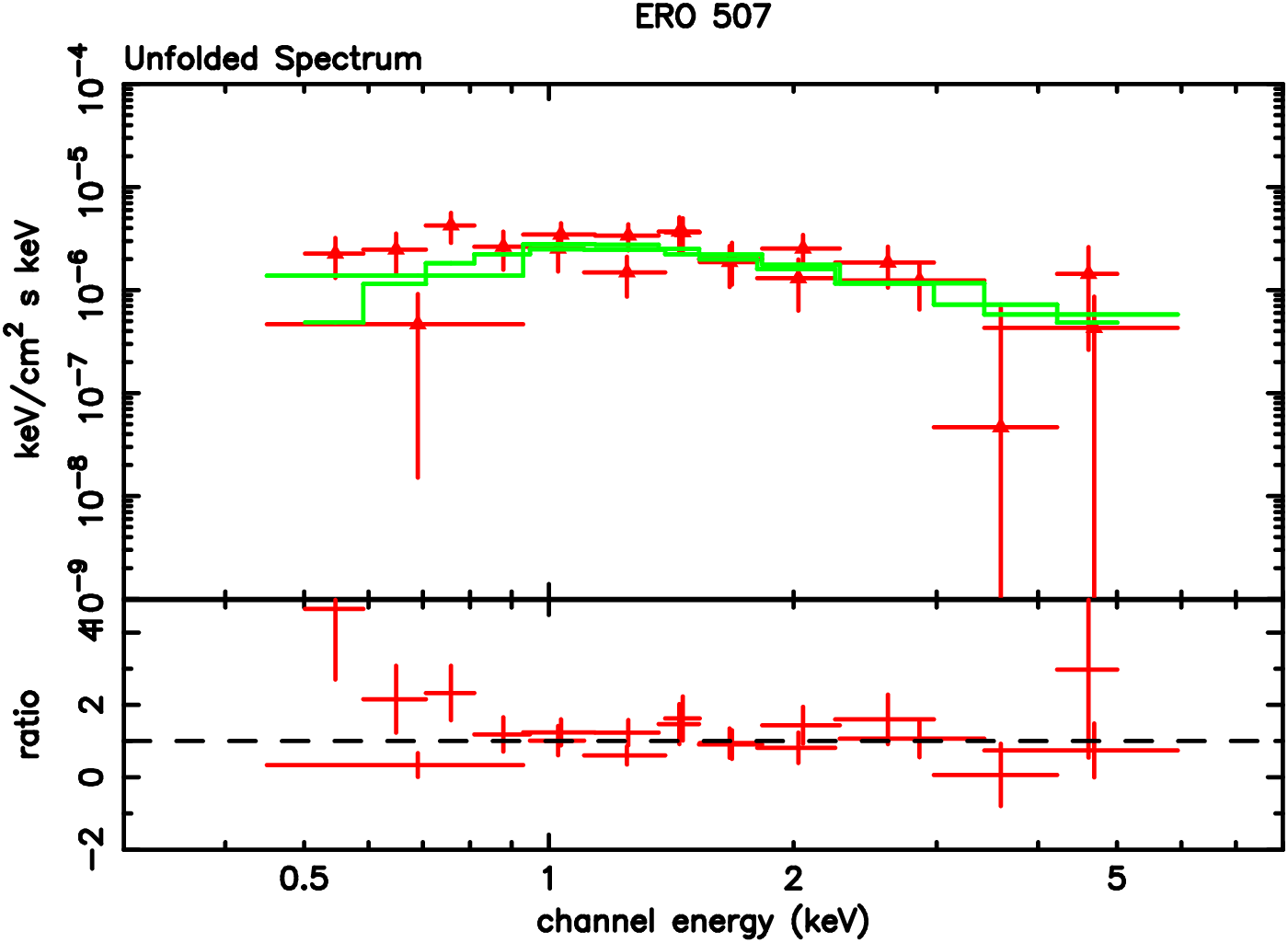}
\includegraphics[width=4cm, height=5cm, angle=-90]{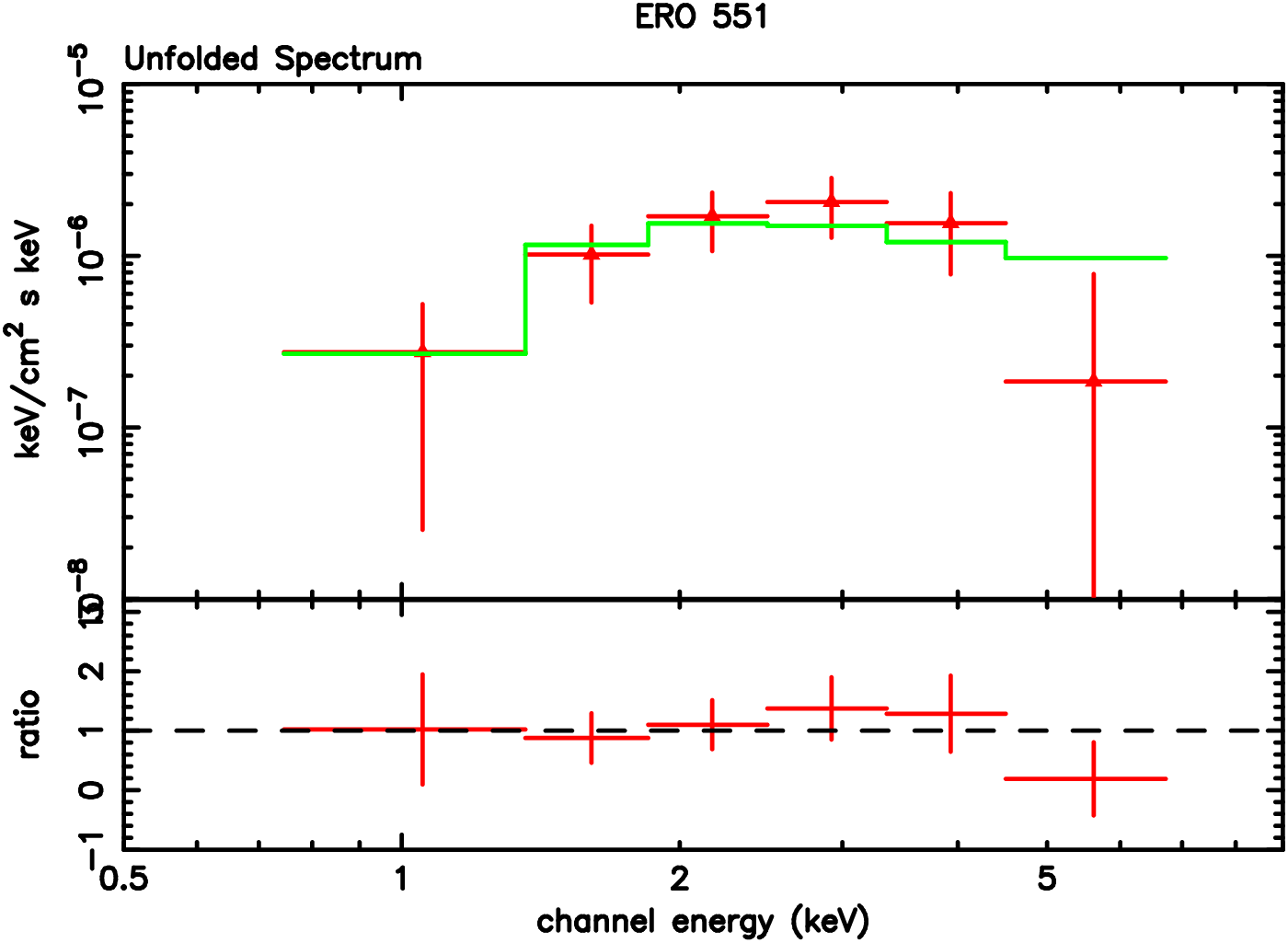}
\end{center}
\caption[]{{\it XMM-Newton} spectra in energy units
(solid points) and best--fit model (continuous line).
Ratios between data and the best--fit model values as a
function of energy are reported in the lower panels.}
\label{eps1}
\end{figure}
\vskip -0.2truecm
A pure thermal component is rejected at more than 97\% confidence level and the
addition of a thermal component to the power-law model is not statistically required.
These results, combined with the best fit parameters obtained, clearly indicate
the presence of high luminosity (unabsorbed L$_{(2-10~keV)}>$10$^{44}$ erg s$^{-1}$), 
obscured (N$_H>$10$^{22}$ cm$^{-2}$) AGN, i.e. QSO2 candidates.

%

\end{document}